\documentclass[prd,twocolumn,aps,psfig,nofootinbib,nobibnotes,superscriptaddress,preprintnumbers,times]{revtex4}
\usepackage{graphicx,bm,color}
\graphicspath{{./fig/}{./png/}}

\newcommand{\BoldVec}[1]{\mathchoice%
  {\mbox{\boldmath $\displaystyle     #1$}}%
  {\mbox{\boldmath $\textstyle        #1$}}%
  {\mbox{\boldmath $\scriptstyle      #1$}}%
  {\mbox{\boldmath $\scriptscriptstyle#1$}}%
}
\newcommand{\EQ}{\begin{equation}}
\newcommand{\EN}{\end{equation}}
\newcommand{\EQA}{\begin{eqnarray}}
\newcommand{\ENA}{\end{eqnarray}}

\newcommand{\Fig}[1]{Fig.~\ref{#1}}

\newcommand{\Tab}[1]{Table~\ref{#1}}
\newcommand{\Figs}[2]{Figs.~\ref{#1} and \ref{#2}}

{}
{}
{}

\newcommand{\tildeh}{\tilde{h}}
\newcommand{\tildeT}{\tilde{T}}

\newcommand{\TTT}{{\sf T}}

\newcommand{\uu}{\BoldVec{u} {}}

\newcommand{\BB}{\BoldVec{B} {}}

\newcommand{\AAA}{\BoldVec{A} {}}

\newcommand{\JJ}{\BoldVec{J} {}}

\newcommand{\kk}{\BoldVec{k} {}}

\newcommand{\nab}{\BoldVec{\nabla} {}}

\newcommand{\SSSS}{\bm{\mathsf{S}}}

\newcommand{\FFF}{\mbox{\boldmath ${\cal F}$} {}}

\newcommand{\dd}{{\rm d} {}}

\def\Imag{\mbox{\rm Im}}

\def\EEM{{\cal E}_{\rm M}}

\def\EEGW{{\cal E}_{\rm GW}}

\def\OmGW{{\Omega}_{\rm GW}}
\def\hrms{{h}_{\rm rms}}

\def\kf{k_{\rm f}}

\def\Brms{B_{\rm rms}}

\def\half{{\textstyle{1\over2}}}

\def\onethird{{\textstyle{1\over3}}}

\newcommand{\neff}{N_{\rm eff}}
\newcommand{\dneff}{\Delta N_{\rm eff}}
\newcommand{\neffv}{N_{\rm eff}^{(\nu)}}

\usepackage{braket}

\begin{document}

\title{Big Bang Nucleosynthesis Limits and Relic Gravitational Waves Detection Prospects}

\date{\today}
\preprint{NORDITA-2021-089}

\author{Tina~Kahniashvili}
\email{tinatin@andrew.cmu.edu}
\affiliation{McWilliams Center for Cosmology and Department of Physics, Carnegie Mellon University, Pittsburgh, PA 15213, USA}
\affiliation{School of Natural Sciences and Medicine, Ilia State University, 0194 Tbilisi, Georgia}
\affiliation{Abastumani Astrophysical Observatory, Tbilisi, GE-0179, Georgia}

\author{Emma Clarke}
\email{emmaclar@andrew.cmu.edu}
\affiliation{McWilliams Center for Cosmology and Department of Physics, Carnegie Mellon University, Pittsburgh, PA 15213, USA}

\author{Jonathan Stepp}
\email{jdstepp@andrew.cmu.edu}
\affiliation{McWilliams Center for Cosmology and Department of Physics, Carnegie Mellon University, Pittsburgh, PA 15213, USA}

\author{Axel~Brandenburg}
\email{brandenb@nordita.org}
\affiliation{Nordita, KTH Royal Institute of Technology and Stockholm University, 10691 Stockholm, Sweden}
\affiliation{The Oskar Klein Centre, Department of Astronomy, Stockholm University, AlbaNova, SE-10691 Stockholm, Sweden}
\affiliation{School of Natural Sciences and Medicine, Ilia State University, 0194 Tbilisi, Georgia}
\affiliation{McWilliams Center for Cosmology and Department of Physics, Carnegie Mellon University, Pittsburgh, PA 15213, USA}

\begin{abstract}
We revisit the big bang nucleosynthesis (BBN) limits on primordial
magnetic fields and/or turbulent motions accounting for the decaying
nature of turbulent sources between the time of generation and BBN.
This leads to larger estimates for the gravitational wave (GW)
signal than previously expected.
We address the detection prospects through space-based interferometers
and pulsar timing arrays or astrometric missions for GWs generated around
the electroweak and quantum chromodynamics energy scale, respectively.
\end{abstract}

\maketitle

Gravitational radiation from the early universe propagates almost freely
throughout the universe's expansion and primordial gravitational waves
(GWs) reflect a precise picture of the very early universe.
Detection of these GWs is a promising tool that would open new avenues
to understand physical processes at energy scales inaccessible
to high energy particle physics experiments but accessible to astrophysical
observation \cite{Caprini:2018mtu}.

There are several milestones of modern cosmology, proven
through cosmic microwave background (CMB) anisotropies and large scale structure statistics.
In particular, the light element abundances allow us to reconstruct the
picture of big bang nucleosynthesis (BBN), but leave several puzzles
prior to BBN (matter-antimatter asymmetry, dynamics of the
universe at the very beginning, nature of dark matter, etc) unsolved.
These unknowns will be reflected in the variety of relic
GW characteristics, including not only the strength of the
signal and its spectral shape, but also its polarization.
Indeed detection of GW polarization is a unique tool to test
fundamental symmetries at these extremely high energies.
If GWs originated from
parity violating sources in the early universe,
they will be circularly polarized and, unlike the CMB, GW
polarization will exist at the basic background and {\it not just the perturbation} level;
see Ref.~\cite{Kahniashvili:2005qi} for pioneering work and see
Refs.~\cite{Alexander:2018fjp,Anand:2018mgf,Niksa:2018ofa,Ellis:2020uid,Kahniashvili:2020jgm,Brandenburg:2021aln}
for recent studies.
This is analogous to the GWs produced via Chern-Simons
coupling \cite{Alexander:2004us,Lyth:2005jf}.
If detected, the GW polarization can be a {\it direct}
measure of the deviations from the standard model (SM)
\cite{Lue:1998mq,Alexander:2009tp,Bartolo:2014hwa}.
One of the major goals of this Letter is to determine whether these
circularly polarized GWs and their polarization are potentially
detectable in the upcoming early-universe GW observation missions
\cite{Amaro-Seoane:2012vvq}.
The strategy to detect the stochastic GW polarization is based
on anisotropy \cite{Seto:2007tn} induced, for example, through our proper motion \cite{Domcke:2019zls,Pol:2021uol}.
Despite promising detection prospects for stochastic GWs through pulsar
timing arrays (PTAs), which are potentially sensitive to GWs generated
around the quantum chromodynamic (QCD) energy scale, detection of the polarization degree
remains problematic.

BBN data (based on light element abundances\footnote{In what follows we
neglect the effects of the strong primordial magnetic field on BBN dynamics as
discussed in Refs.~\cite{PhysRevD.104.123534,Lu:2022aus}}) impose an upper limit
on the universe's expansion rate, e.g., the Hubble parameter,
$H = d {\rm ln} a/dt_{\rm phys}$ (with physical time $t_{\rm phys}$
and scale factor $a$), and
correspondingly, on additional relativistic species such as massless
(or ultrarelativistic) hypothetical particles, early stage dark energy
(or any bosonic massless field), dark radiation, electromagnetic
fields or early-universe plasma motions (turbulence), relic GWs, etc
\cite{Frieman:1989fx, Grasso:1994ph, Fuller:1996hx, Ichikawa:2004ju,
Simha:2008zj, Sasankan:2017eqr, Keith:2020jww}.
Conventionally, the energy density of these additional relativistic
components is characterized in terms of the {\it effective number of
relativistic species}, $N_{\rm eff}$.
The SM predicts for neutrino species
$\neffv = 3.046$, which is slightly larger than 3 because
neutrinos did not decouple instantaneously and were still able to
interact with electrons and positrons near electron-positron annihilation
\cite{deSalas:2016ztq}.
Other additional relativistic
components contribute $\Delta \neff = \neff - \neffv$ to this
{\it effective} neutrino count.
Notably, the presence of additional relativistic components does not spoil
the time dependence of the scale factor during the radiation-dominated
epoch, but it does affect the Hubble parameter and Hubble time scale, $H^{-1}$.
The joint analysis of CMB measurements and BBN light
element abundances put $\neff = 2.862 \pm 0.306$ at 95\% confidence
\cite{Fields:2019}.
Using the upper bound of this error interval ($\neff = 3.168$),
we express the maximum ratio of additional components
of energy density \(\rho_\mathrm{add} \) to the radiation
energy density \( \rho_\mathrm{rad} \) at the BBN temperature
as $\rho_\mathrm{add}/\rho_\mathrm{rad} \simeq 0.0277
\left(\dneff/0.122 \right)$, normalized around $\dneff = 0.122$.
The maximum value of this ratio is limited by the
combined CMB and BBN data.
We note that this upper bound coincides with the constraint
on the GW contribution to the radiation energy density found in
Ref.~\cite{Benetti:2021uea} using CMB and BBN data combined with limits
from NANOGrav and late-time measurements of the expansion history.
Interestingly, the light element abundances (with the bounds on $\neff$)
impose limits on the lepton asymmetry in the universe \cite{Simha:2008mt}
that might result in primordial chiral magnetic fields \cite{PhysRevD.22.3080}
and correspondingly serve as a source for polarized GWs \cite{Brandenburg:2021aln}.

In this Letter we address the BBN bounds from the point of view of
early-universe anisotropic stress (namely primordial magnetic fields and
turbulent sources) and the induced GW signal.
Inhomogeneous magnetic fields are known to affect the primordial
lithium abundance \cite{Luo:2018nth}.
Here, however, we are particularly interested in the strength, the
spectral shape, and the polarization degree of the induced GWs.
Violent processes in the early universe might lead to the
development of turbulence.
In particular, first order electroweak (EW) and QCD phase-transition bubble collisions and nucleation might lead
to turbulent plasma motions \cite{Turner:1990rc, Witten:1984rs,
Hogan:1986qda, Kamionkowski:1993fg}, or, alternatively, turbulence can
be induced by primordial magnetic fields coupled to the cosmological plasma
\cite{Brandenburg:1996fc,Christensson:2000sp,Kahniashvili:2010gp,Brandenburg:2017rnt,Durrer:2013pga}.
The stochastic GW background from these turbulent sources has been studied
for decades now; see Refs.~\cite{Kamionkowski:1993fg, Kosowsky:2001xp,
Dolgov:2002ra} for pioneering works and Ref.~\cite{Caprini:2019egz}
for a review and references therein.

Both analytical and numerical studies suggest that a strong enough
gravitational radiation signal (when and if the total energy density of
the source is a substantial fraction of the total (radiation) energy density,
$\rho_{\rm rad}$, at the moment of the GW
generation,
characterized by the temperature, $T_\star$,
and the number of relativistic degrees of freedom, $g_\star$,
where here and below an asterisk denotes the generation moment)
is detectable by space-based missions, such as the Laser Interferometer
Space Antenna (LISA) (for GWs generated around the EW energy
scale) \cite{Caprini:2019egz}, or by PTAs, such as
NANOGrav \cite{NANOGrav:2020bcs}, and astrometric missions such
as GAIA \cite{Garcia-Bellido:2021zgu} (for GWs generated around the
QCD energy scale).
Notably, the NANOGrav collaboration recently announced strong evidence
for a stochastic GW signal \cite{NANOGrav:2020bcs} that might be
associated with primordial sources \cite{NANOGrav:2021flc}.
On the other hand, when estimating the strength of the GW signal, the
maximum allowed source energy density was assumed to be determined by
the BBN bounds discussed above (i.e., not exceeding a few percent of the
total radiation energy, i.e. ${\mathcal E}_{\rm turb} = \zeta \rho_{\rm rad} $
with the parameter $\zeta$ being in general time-dependent and being a few per cent at BBN).

\begin{figure*}[t]
 \centering
 \includegraphics[scale=0.6]{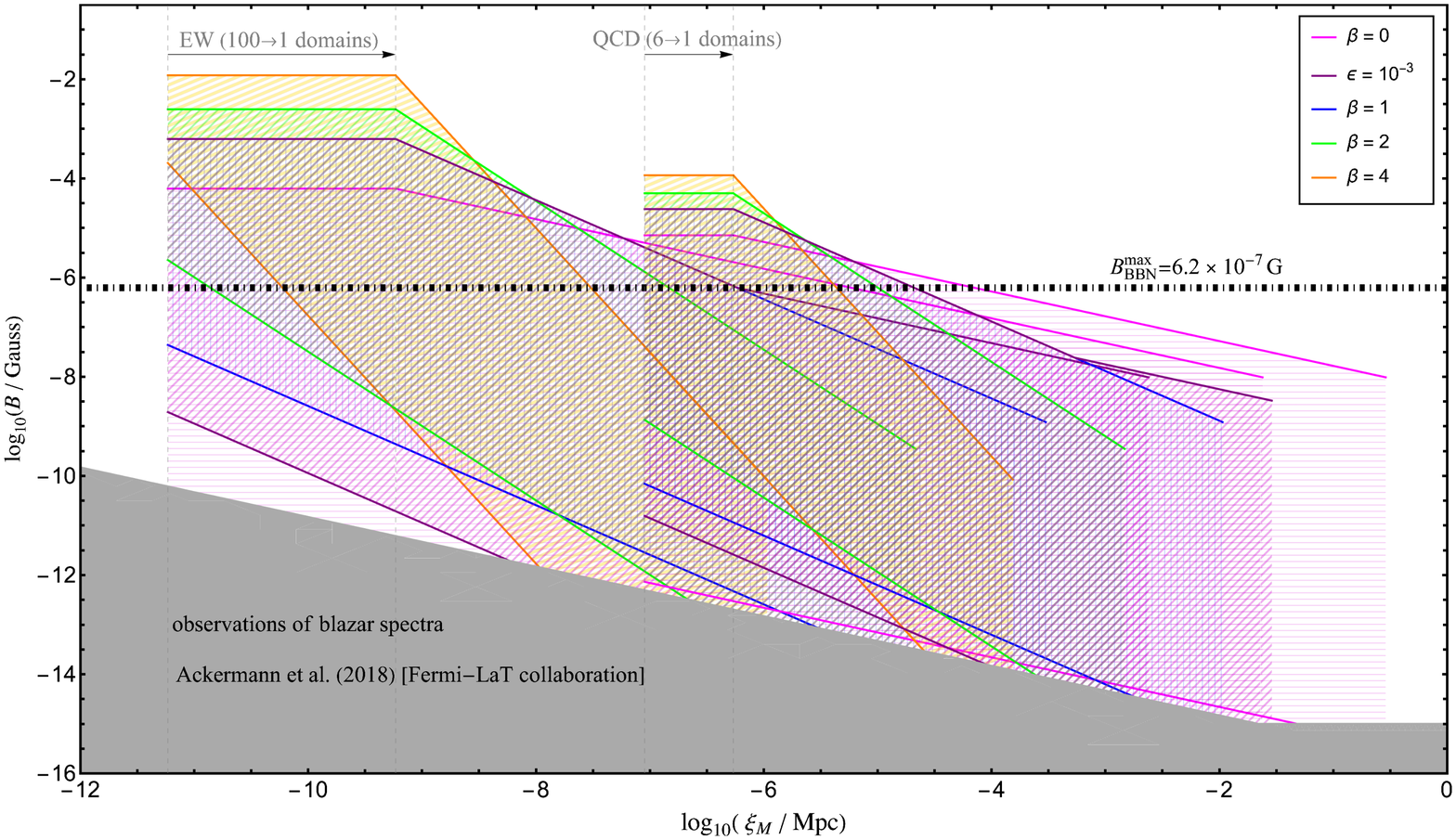}
\caption{
Possible turbulent evolution of the comoving magnetic field strength
$B$ (decaying with time) and correlation length $\xi_M$
(increasing with time) from generation at the EW and QCD scales in
the cases of fully helical ($\beta = 0$), nonhelical ($\beta =$1, 2, 4),
and partially helical (with $\epsilon_{M,\star}=10^{-3}$) MHD turbulence.
Upper limits on the correlation length are determined by the size of the horizon
and number of domains (bubbles) at generation,
ranging from 1 to 6 (at QCD) or 100 (at EW), depending on the phase transition modeling.
Lines terminate (on the right) at recombination ($T=0.25$ eV).
The upper limit of the comoving field strength at BBN ($T$ = 0.1 MeV) is indicated by the black dot-dashed line.
Regimes excluded by observations of blazar spectra \cite{Ackermann_2018} are marked in gray.
The hatched regions correspond to possible trajectories bounded by an
(upper) limit from BBN and a
(lower) limit from the blazar spectra.
}\label{fig:evolution}
\end{figure*}

Due to weak coupling between gravity and matter
(i.e., the smallness of Newton's constant $G$), the GW generation
from any turbulent source is characterized by low efficiency
and, consequently, the ratio $\zeta$ of turbulent source energy density
to the total radiation energy is not affected
by the emission of gravitational radiation.
In other words, the energy radiated in GWs will not induce substantial
damping of the turbulent energy density.
Moreover, if turbulent decay processes are discarded (i.e., velocity and
magnetic fields are ``frozen-in'' to the primordial plasma),
$\zeta$ is unchanged during the radiation-dominated epoch.
Applying this logic to the BBN bounds, the few percent limit was
applied {\it a priori} to much earlier time scales when GWs were generated.
As it was seen in simulations \cite{RoperPol:2019wvy,
Kahniashvili:2020jgm}, the GW energy density reaches a maximum and stays
unchanged after a short time.
Thus, only $\zeta$ at {\it the moment of the source
activation} (i.e., GW generation) matters.

In the case of decaying turbulence, the situation is different:
$\zeta$ is time dependent and the decay rate is determined by
the specific model of turbulence.
Decaying turbulence leads to a power-law decay
${\mathcal E}_{\rm turb}(t) \propto (t/t_\star)^{-p}$, and a
growth of the correlation length $\xi_{\rm turb}$ by an inverse
cascade mechanism such that $\xi_{\rm turb} \propto (t/t_\star)^{q}$, where
$t=\int dt_{\rm phys}/a$ is the conformal time and the parameters $p$ and $q$
depend on the properties of the turbulence (e.g., in helical turbulence
$p=q=2/3$, while for non-helical magnetically dominated turbulence $p=1$
and $q=1/2$, but other variants are possible).
The scaling exponent $q$ may reflect the presence of an underlying
conservation law (helicity conservation, Loitsiansky integral) and is
also determined by the nature of turbulence (kinetically or magnetically
dominated).
The combined values of $p$ and $q$ for a particular process can
be summarized by the parameter $\beta=p/q-1$, which characterizes the
decay of the spectral peak of magnetic energy \cite{Brandenburg:2017neh}.
Partially helical magnetic fields are also described by their
fractional helicity, i.e., the ratio of the magnetic helicity to its maximal
value, $\epsilon_{M,\star} < 1$.
Due to this decay, the BBN bound allows larger values of $\zeta$ at the
moment of GW generation, making the GW signal stronger.
The maximum allowed energy density of turbulent sources that satisfy
the BBN limits will be different at the EW and QCD energy scales
(EW turbulence has a longer decay period, allowing higher
values for the initial energy density that still satisfies the BBN bounds).

Figure~\ref{fig:evolution} shows the bounds on the strength of the
magnetic fields at their generation (EW or QCD scales) determined such
that the strength does not exceed the upper limit of the comoving field
strength at BBN \cite{SM1} and is above the lower observational bounds
on the field strength at recombination.
We see that allowed values for the
magnetic fields at the moment of generation (upper left end of the
lines) are not constrained to microGauss field strength, as it was
claimed previously based on BBN bounds without accounting for decaying
turbulence \cite{Vachaspati:2020blt}.
In fact, if we were previously considering an Alfv\'en speed $v_A$
or characteristic velocities of 0.2--0.3 (in units
of the speed of light), the new limits possibly imply $v_A
\rightarrow 1$ \cite{RoperPol:2019wvy}.
Obviously, in this case we deal with relativistic turbulence that might
be characterized by different decay laws or efficiency to generate GWs.
However, recent relativistic turbulence numerical simulations
\cite{Zrake:2015hda} show that the basic properties of turbulence decay
are preserved, including non-helical inverse cascading.
Also, following arguments of Ref.~\cite{Kosowsky:2001xp}, the
non-relativistic description of turbulent sources results in
an underestimation of the signal.

\begin{figure*}[t!]\begin{center}
\includegraphics[width=\textwidth]{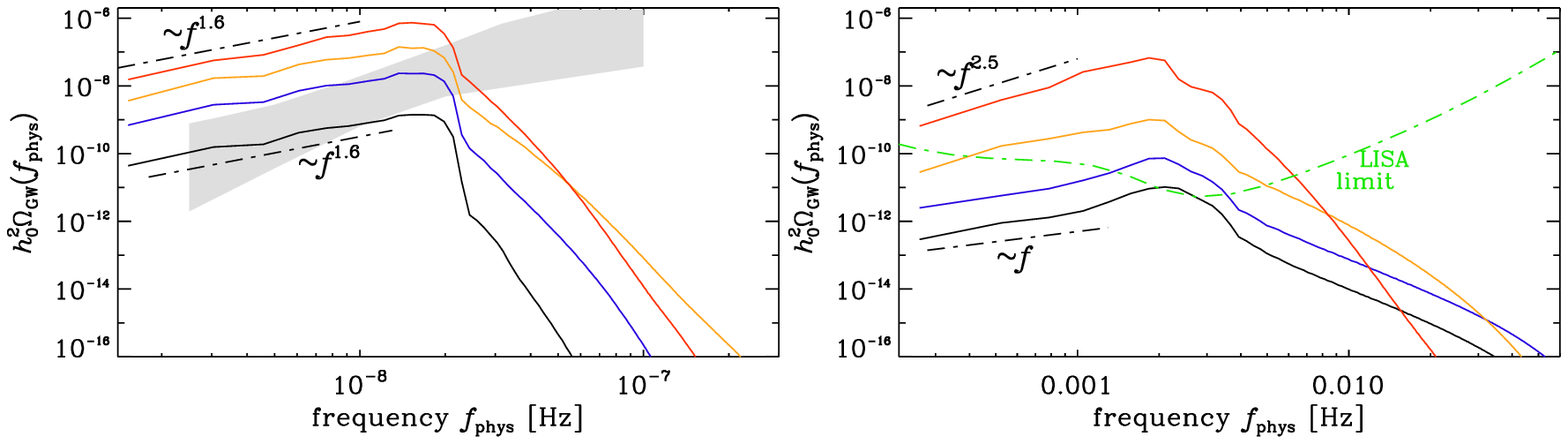}
\end{center}\caption[]{
GW energy spectra (per logarithmic frequency interval), $h_0^2\OmGW(f)$,
for both the QCD Runs~a--d (left) and the EW Runs~A--D (right) scales
shown in red, orange, blue, and black, respectively.
}\label{pspecm_wkf6}\end{figure*}

Below we present the first simulations of the GW signal from such strong turbulence sources.
We use the {\sc Pencil Code} \cite{JOSS,PC} to simulate magnetohydrodynamic (MHD) turbulence in the early universe
by computing the stochastic GW background and relic magnetic fields \cite{RoperPol:2019wvy}.
Turbulence is driven by applying an electromagnetic force that is
$\delta$-correlated in time and has the desired spatial spectrum.
We vary the forcing strength and adjust the viscosity such that the smallest length
scales in the simulation are sufficiently well resolved to dissipate the injected
energy near the highest available wavenumber.
We perform runs for the QCD and EW energy scales; see Ref.~\cite{SM1}
for a table summarizing the eight runs presented in this Letter.

\begin{figure*}[t!]\begin{center}
\includegraphics[width=\textwidth]{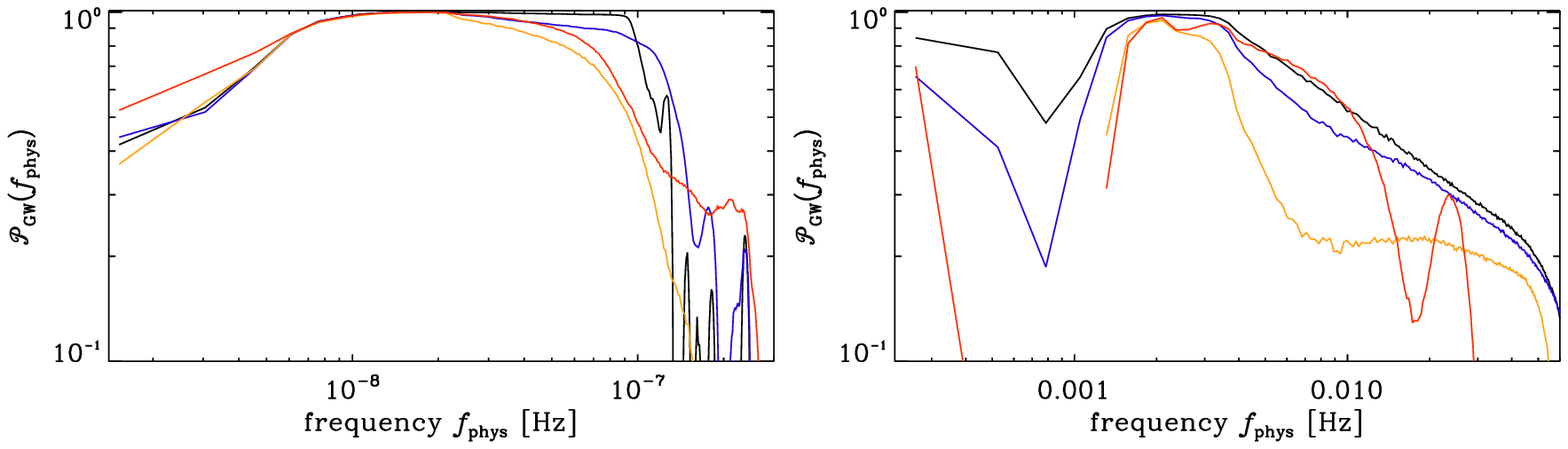}
\end{center}\caption[]{
Polarization spectra, ${\cal P}_{\rm GW}(f)$, for the
QCD Runs~a--d (left) and the EW Runs~A--D (right) scales \cite{SM1}
shown in red, orange, blue, and black, respectively.
}\label{pspecm_wkf6_600_pol}\end{figure*}

The GW detection prospects are strongly affected by the characteristic
frequency ranges and thus the energy-containing wave number of the source.
More precisely, the GW spectrum peaks at the comoving angular frequency
$\omega_{\rm peak} =(2\pi f_{\rm peak}) =2 k_0$,
where $k_0$ is the initial peak
wave number of the source energy density spectrum (in natural units $c=1$).
The inertial wave number is determined by the turbulent
eddy size ($k_0 = 2\pi/L$), and if we assume that turbulence arises from
phase transitions, the eddy size may be associated with the bubble size
\cite{Kahniashvili:2009mf}.
Independently of the nature the turbulence, the typical
length scale is limited by the Hubble scale.
In what follows, we use the characteristic wave number $k_0$
normalized by the Hubble wave number $H_\star$.

The energy density of early-universe turbulent sources is determined by
the efficiency of converting the available radiation energy into
turbulent energy.
In the case of first-order phase transitions, it
can be expressed in the terms of the parameter
$\alpha = \rho_{\rm vac}/\rho_{\rm rad}=4\rho_{\rm vac}/3(\rho+P)$
(with $\rho$ and $P$ being the plasma energy density and pressure, respectively) --
the ratio between the latent heat (false vacuum energy) density and the
plasma radiation energy density (which is determined at the phase transition
temperature \cite{Kamionkowski:1993fg}).
$\alpha \sim$
a few corresponds to extremely strong phase transitions.
Ref.~\cite{Ellis:2018mja} discusses a few beyond-SM models, which could
include first-order phase transitions, and some of these models predict
$\alpha \gtrsim 1$ for specific ranges of their parameter spaces.
In particular, the addition of a 6-dimensional term to the Higgs
potential \cite{Bodeker:2004ws} or the addition of a singlet scalar field
\cite{Choi:1993cv} allow for these particularly strong phase transitions.
The induced turbulence can then be characterized by the velocity
$v_i = 1/\sqrt{1+(\rho+P)/{(2\mathcal{E}_i})}$,
which refers either to the turbulent velocity $v_T$ or the
effective Alfv\'en velocity $v_A$, and $\mathcal{E}_i$ refers to either
the kinetic, $\mathcal{E}_K$, or magnetic, $\mathcal{E}_M$, energy density.
By defining the efficiency coefficient $\kappa \equiv \kappa(\alpha) \in (0,1)$
(which increases with $\alpha$), i.e., the fraction of vacuum energy
that is transformed into $\mathcal{E}_K$ or $\mathcal{E}_M$,
rather than into heat \cite{Kosowsky:2001xp},
we recover {\it relativistic expressions}
for turbulent motions, $v_T=1/\sqrt{1+{4}/{(3\kappa \alpha)}}$
\cite{Nicolis:2003tg}, and the Alfv\'en velocity,
$v_A =1/\sqrt{1+(4/3)\rho/(2\mathcal{E}_M})$
\cite{Gedalin:1993},\footnote{In the non-relativistic
limit we obtain
$v_T = \sqrt{2{\mathcal{E}_K}/(\rho+P)}$ and
$v_A = B_{\rm eff}/\sqrt {4\pi (\rho+P)}=\sqrt{2{\mathcal{E}_M}/(\rho+P)}$}
while previous studies (see Ref.~\cite{Caprini:2019egz}
for a review and references therein)
assumed non-relativistic motions.

The additional relativistic degrees of freedom in the early universe due
to the addition of the energy densities of the turbulent sources can
be subsumed into $\dneff$.
This increase in $\neff$ increases the CMB-inferred value of the
Hubble constant, $H_0$, helping to reduce the tension with
late-universe values.
A value of $\dneff \sim 0.4$ could alleviate the Hubble tension
\cite{Riess:2016jrr}.
Interestingly, it has been shown that the recent NANOGrav results may
also favor a larger value of $\neff$ \cite{Nakai:2020oit} if the signal
arises in the early universe.
Even though large values of $\alpha$ are not restricted by currently available
BBN or other observational data, we limit ourselves by $\alpha_{PT} \leq 1$
that was addressed previously in several studies, see Ref.~\cite{Caprini:2018mtu} and references therein.

In Figure \ref{pspecm_wkf6}, we present GW spectra (per logarithmic frequency
interval and normalized by the critical energy density) from our
simulations expressed as $h_0^2\Omega_{\rm GW}(f_{\rm phys})$ for two
families of models considered previously: one for the EW scale with
$k/H_\star=600$ \cite{RoperPol:2019wvy} and one for the QCD phase
transition with $k/H_\star=6$ \cite{Brandenburg:2021tmp}.
The former set of models is similar to simulations of
Ref.~\cite{Kahniashvili:2020jgm}, except that now we also consider
models with stronger turbulent driving which is applied over one Hubble
time along with a period during which the forcing decreases linearly
in time to zero, again over one Hubble time.

As already noted in previous studies \cite{RoperPol:2019wvy,
Kahniashvili:2020jgm, Brandenburg:2021aln}, the GW energy spectrum from
forced turbulence shows a rapidly declining inertial range for
frequencies above the peak.
This is because only the smallest wave numbers contribute significantly
to the driving of GWs \cite{RoperPol:2019wvy,Brandenburg:2021bvg}.
The GW energy $h_0^2\Omega_{\rm GW}(f_{\rm phys})$ scales
approximately quadratically with the ratio of magnetic energy to
characteristic wave number $k_0$ as $(Q\EEM/k_0)^2$, where $Q$ is the GW
efficiency (of order unity).
For the QCD phase transition, the characteristic wave number is a hundred
times smaller, so the GW energy is correspondingly larger.

Toward smaller frequencies, the spectra show a shallower fall-off,
in some cases proportional to $f_{\rm phys}^{1.6}$.
This is steeper than what has been found in earlier simulations
at lower magnetic energies, but
shallower than what was generally expected based on analytical considerations.
Physics beyond
the SM often leads to parity symmetry breaking
and correspondingly to polarized gravitational waves.
In Figure \ref{pspecm_wkf6_600_pol} we show the polarization spectra,
${\cal P}_{\rm GW}(k) = \!\int\!2\,\Imag\,\tildeh_+\tildeh_\times^*\,k^2\dd\Omega_k
/\!\!\int\!(|\tildeh_+|^2+|\tildeh_\times|^2) k^2\dd\Omega_k$,
for the same runs as in 
Figure \ref{pspecm_wkf6};
see also Eq.~(B.17) in Ref.~\cite{RoperPol:2018sap}.
For the QCD phase transition with only a few bubbles per linear
Hubble scale, the polarization spectra have an extended region with
${\mathcal P}_{\rm GW} \sim 1$, while for the electroweak phase transitions with
tens of bubbles, the polarization spectra have non-trivial profiles
with a narrower plateau.

In summary, BBN data do not limit the kinetic or magnetic energy density
of the turbulence at the moment of its
generation to be 10\% of the radiation energy when the decay process
is accounted for.
Strong turbulence unavoidably results in a more powerful source for
the GW signal with more optimistic prospects for GW detection.

\vspace{2mm}
{\bf Data availability}---The source code used for the
simulations of this study, the {\sc Pencil Code},
is freely available from Refs.~\cite{JOSS,PC}.
The simulation setups and the corresponding data
are freely available from Ref.~\cite{DATA}.

\vspace{2mm}
{\bf Acknowledgements}---Support through the Swedish Research Council, grant 2019-04234,
and Shota Rustaveli NSF of Georgia (grant FR/18-1462) are gratefully acknowledged.
Nordita is supported in part by Nordforsk.
We acknowledge the allocation of computing resources provided by the
Swedish National Infrastructure for Computing (SNIC)
at the PDC Center for High Performance Computing Stockholm
and the National Supercomputer Centre (NSC) at Link\"oping.
J.S. acknowledges
support from the Undergraduate Research Office in the form of a Summer Undergraduate Research Fellowship.

\bibliographystyle{h-physrev5}
\bibliography{GWs-BBN}

\begin{thebibliography}{10}

\bibitem{Caprini:2018mtu}
C.~Caprini and D.~G. Figueroa,
\newblock Class. Quant. Grav. {\bf 35}, 163001 (2018), arXiv:1801.04268.

\bibitem{Kahniashvili:2005qi}
T.~Kahniashvili, G.~Gogoberidze, and B.~Ratra,
\newblock Phys. Rev. Lett. {\bf 95}, 151301 (2005), arXiv:astro-ph/0505628.

\bibitem{Alexander:2018fjp}
S.~Alexander, E.~McDonough, and D.~N. Spergel,
\newblock JCAP {\bf 05}, 003 (2018), arXiv:1801.07255.

\bibitem{Anand:2018mgf}
S.~Anand, J.~R. Bhatt, and A.~K. Pandey,
\newblock Eur. Phys. J. C {\bf 79}, 119 (2019), arXiv:1801.00650.

\bibitem{Niksa:2018ofa}
P.~Niksa, M.~Schlederer, and G.~Sigl,
\newblock Class. Quant. Grav. {\bf 35}, 144001 (2018), arXiv:1803.02271.

\bibitem{Ellis:2020uid}
J.~Ellis, M.~Fairbairn, M.~Lewicki, V.~Vaskonen, and A.~Wickens,
\newblock JCAP {\bf 10}, 032 (2020), arXiv:2005.05278.

\bibitem{Kahniashvili:2020jgm}
T.~Kahniashvili, A.~Brandenburg, G.~Gogoberidze, S.~Mandal, and A.~Roper~Pol,
\newblock Phys. Rev. Res. {\bf 3}, 013193 (2021), arXiv:2011.05556.

\bibitem{Brandenburg:2021aln}
A.~Brandenburg, Y.~He, T.~Kahniashvili, M.~Rheinhardt, and J.~Schober,
\newblock Astrophys. J. {\bf 911}, 110 (2021), arXiv:2101.08178.

\bibitem{Alexander:2004us}
S.~H.-S. Alexander, M.~E. Peskin, and M.~M. Sheikh-Jabbari,
\newblock Phys. Rev. Lett. {\bf 96}, 081301 (2006), arXiv:hep-th/0403069.

\bibitem{Lyth:2005jf}
D.~H. Lyth, C.~Quimbay, and Y.~Rodriguez,
\newblock JHEP {\bf 03}, 016 (2005), arXiv:hep-th/0501153.

\bibitem{Lue:1998mq}
A.~Lue, L.-M. Wang, and M.~Kamionkowski,
\newblock Phys. Rev. Lett. {\bf 83}, 1506 (1999), arXiv:astro-ph/9812088.

\bibitem{Alexander:2009tp}
S.~Alexander and N.~Yunes,
\newblock Phys. Rept. {\bf 480}, 1 (2009), arXiv:0907.2562.

\bibitem{Bartolo:2014hwa}
N.~Bartolo, S.~Matarrese, M.~Peloso, and M.~Shiraishi,
\newblock JCAP {\bf 01}, 027 (2015), arXiv:1411.2521.

\bibitem{Amaro-Seoane:2012vvq}
P.~Amaro-Seoane {\em et~al.},
\newblock Class. Quant. Grav. {\bf 29}, 124016 (2012), arXiv:1202.0839.

\bibitem{Seto:2007tn}
N.~Seto and A.~Taruya,
\newblock Phys. Rev. Lett. {\bf 99}, 121101 (2007), arXiv:0707.0535.

\bibitem{Domcke:2019zls}
V.~Domcke {\em et~al.},
\newblock JCAP {\bf 05}, 028 (2020), arXiv:1910.08052.

\bibitem{Pol:2021uol}
A.~Roper~Pol, S.~Mandal, A.~Brandenburg, and T.~Kahniashvili,
\newblock (2021), arXiv:2107.05356.

\bibitem{PhysRevD.104.123534}
M.~Kusakabe, A.~Kedia, G.~J. Mathews, and N.~Sasankan,
\newblock Phys. Rev. D {\bf 104}, 123534 (2021).

\bibitem{Lu:2022aus}
Y.~Lu and M.~Kusakabe,
\newblock Astrophys. J. Lett. {\bf 926}, L4 (2022), arXiv:2201.13039.

\bibitem{Frieman:1989fx}
J.~A. Frieman, E.~W. Kolb, and M.~S. Turner,
\newblock Phys. Rev. D {\bf 41}, 3080 (1990).

\bibitem{Grasso:1994ph}
D.~Grasso and H.~R. Rubinstein,
\newblock Astropart. Phys. {\bf 3}, 95 (1995), arXiv:astro-ph/9409010.

\bibitem{Fuller:1996hx}
G.~M. Fuller and C.~Y. Cardall,
\newblock Nucl. Phys. B Proc. Suppl. {\bf 51}, 71 (1996),
  arXiv:astro-ph/9606025.

\bibitem{Ichikawa:2004ju}
K.~Ichikawa and M.~Kawasaki,
\newblock Phys. Rev. D {\bf 69}, 123506 (2004), arXiv:hep-ph/0401231.

\bibitem{Simha:2008zj}
V.~Simha and G.~Steigman,
\newblock JCAP {\bf 06}, 016 (2008), arXiv:0803.3465.

\bibitem{Sasankan:2017eqr}
N.~Sasankan, M.~R. Gangopadhyay, G.~J. Mathews, and M.~Kusakabe,
\newblock Int. J. Mod. Phys. E {\bf 26}, 1741007 (2017), arXiv:1706.03630.

\bibitem{Keith:2020jww}
C.~Keith, D.~Hooper, N.~Blinov, and S.~D. McDermott,
\newblock Phys. Rev. D {\bf 102}, 103512 (2020), arXiv:2006.03608.

\bibitem{deSalas:2016ztq}
P.~F. de~Salas and S.~Pastor,
\newblock JCAP {\bf 07}, 051 (2016), arXiv:1606.06986.

\bibitem{Fields:2019}
B.~D. {Fields}, K.~A. {Olive}, T.-H. {Yeh}, and C.~{Young},
\newblock Journal of Cosmology and Astroparticle Physics {\bf 2020}, 010
  (2020), arXiv:1912.01132.

\bibitem{Benetti:2021uea}
M.~Benetti, L.~L. Graef, and S.~Vagnozzi,
\newblock (2021), arXiv:2111.04758.

\bibitem{Simha:2008mt}
V.~Simha and G.~Steigman,
\newblock JCAP {\bf 08}, 011 (2008), arXiv:0806.0179.

\bibitem{PhysRevD.22.3080}
A.~Vilenkin,
\newblock Phys. Rev. D {\bf 22}, 3080 (1980).

\bibitem{Luo:2018nth}
Y.~Luo, T.~Kajino, M.~Kusakabe, and G.~J. Mathews,
\newblock Astrophys. J. {\bf 872}, 172 (2019), arXiv:1810.08803.

\bibitem{Turner:1990rc}
M.~S. Turner and F.~Wilczek,
\newblock Phys. Rev. Lett. {\bf 65}, 3080 (1990).

\bibitem{Witten:1984rs}
E.~Witten,
\newblock Phys. Rev. D {\bf 30}, 272 (1984).

\bibitem{Hogan:1986qda}
C.~J. Hogan,
\newblock Mon. Not. Roy. Astron. Soc. {\bf 218}, 629 (1986).

\bibitem{Kamionkowski:1993fg}
M.~Kamionkowski, A.~Kosowsky, and M.~S. Turner,
\newblock Phys. Rev. D {\bf 49}, 2837 (1994), arXiv:astro-ph/9310044.

\bibitem{Brandenburg:1996fc}
A.~Brandenburg, K.~Enqvist, and P.~Olesen,
\newblock Phys. Rev. D {\bf 54}, 1291 (1996), arXiv:astro-ph/9602031.

\bibitem{Christensson:2000sp}
M.~Christensson, M.~Hindmarsh, and A.~Brandenburg,
\newblock Phys. Rev. E {\bf 64}, 056405 (2001), arXiv:astro-ph/0011321.

\bibitem{Kahniashvili:2010gp}
T.~Kahniashvili, A.~Brandenburg, A.~G. Tevzadze, and B.~Ratra,
\newblock Phys. Rev. D {\bf 81}, 123002 (2010), arXiv:1004.3084.

\bibitem{Brandenburg:2017rnt}
A.~Brandenburg {\em et~al.},
\newblock Phys. Rev. Fluids. {\bf 4}, 024608 (2019), arXiv:1710.01628.

\bibitem{Durrer:2013pga}
R.~Durrer and A.~Neronov,
\newblock Astron. Astrophys. Rev. {\bf 21}, 62 (2013), arXiv:1303.7121.

\bibitem{Kosowsky:2001xp}
A.~Kosowsky, A.~Mack, and T.~Kahniashvili,
\newblock Phys. Rev. D {\bf 66}, 024030 (2002), arXiv:astro-ph/0111483.

\bibitem{Dolgov:2002ra}
A.~D. Dolgov, D.~Grasso, and A.~Nicolis,
\newblock Phys. Rev. D {\bf 66}, 103505 (2002), arXiv:astro-ph/0206461.

\bibitem{Caprini:2019egz}
C.~Caprini {\em et~al.},
\newblock JCAP {\bf 03}, 024 (2020), arXiv:1910.13125.

\bibitem{NANOGrav:2020bcs}
NANOGrav, Z.~Arzoumanian {\em et~al.},
\newblock Astrophys. J. Lett. {\bf 905}, L34 (2020), arXiv:2009.04496.

\bibitem{Garcia-Bellido:2021zgu}
J.~Garcia-Bellido, H.~Murayama, and G.~White,
\newblock (2021), arXiv:2104.04778.

\bibitem{NANOGrav:2021flc}
NANOGrav, Z.~Arzoumanian {\em et~al.},
\newblock (2021), arXiv:2104.13930.

\bibitem{Ackermann_2018}
M.~Ackermann {\em et~al.},
\newblock The Astrophysical Journal Supplement Series {\bf 237}, 32 (2018).

\bibitem{RoperPol:2019wvy}
A.~Roper~Pol, S.~Mandal, A.~Brandenburg, T.~Kahniashvili, and A.~Kosowsky,
\newblock Phys. Rev. D {\bf 102}, 083512 (2020), arXiv:1903.08585.

\bibitem{Brandenburg:2017neh}
A.~Brandenburg {\em et~al.},
\newblock Phys. Rev. D {\bf 96}, 123528 (2017), arXiv:1711.03804.

\bibitem{SM1}
See the supplemental material for the full set of equations and a summary of
  additional parameters of the simulations.

\bibitem{Vachaspati:2020blt}
T.~Vachaspati,
\newblock Rept. Prog. Phys. {\bf 84}, 074901 (2021), arXiv:2010.10525.

\bibitem{Zrake:2015hda}
J.~Zrake and W.~E. East,
\newblock Astrophys. J. {\bf 817}, 89 (2016), arXiv:1509.00461.

\bibitem{JOSS}
{Pencil Code Collaboration} {\em et~al.},
\newblock The Journal of Open Source Software {\bf 6}, 2807 (2021).

\bibitem{PC}
The pencil code. doi:10.5281/zenodo.2315093.
  \href{https://github.com/pencil-code}{https://github.com/pencil-code}.

\bibitem{Kahniashvili:2009mf}
T.~Kahniashvili, L.~Kisslinger, and T.~Stevens,
\newblock Phys. Rev. D {\bf 81}, 023004 (2010), arXiv:0905.0643.

\bibitem{Ellis:2018mja}
J.~Ellis, M.~Lewicki, and J.~M. No,
\newblock JCAP {\bf 04}, 003 (2019), arXiv:1809.08242.

\bibitem{Bodeker:2004ws}
D.~Bodeker, L.~Fromme, S.~J. Huber, and M.~Seniuch,
\newblock JHEP {\bf 02}, 026 (2005), arXiv:hep-ph/0412366.

\bibitem{Choi:1993cv}
J.~Choi and R.~R. Volkas,
\newblock Phys. Lett. B {\bf 317}, 385 (1993), arXiv:hep-ph/9308234.

\bibitem{Nicolis:2003tg}
A.~Nicolis,
\newblock Class. Quant. Grav. {\bf 21}, L27 (2004), arXiv:gr-qc/0303084.

\bibitem{Gedalin:1993}
M.~{Gedalin},
\newblock \pre {\bf 47}, 4354 (1993).

\bibitem{Riess:2016jrr}
A.~G. Riess {\em et~al.},
\newblock Astrophys. J. {\bf 826}, 56 (2016), arXiv:1604.01424.

\bibitem{Nakai:2020oit}
Y.~Nakai, M.~Suzuki, F.~Takahashi, and M.~Yamada,
\newblock Phys. Lett. B {\bf 816}, 136238 (2021), arXiv:2009.09754.

\bibitem{Brandenburg:2021tmp}
A.~Brandenburg, E.~Clarke, Y.~He, and T.~Kahniashvili,
\newblock Phys. Rev. D {\bf 104}, 043513 (2021), arXiv:2102.12428.

\bibitem{Brandenburg:2021bvg}
A.~Brandenburg {\em et~al.},
\newblock Class. Quant. Grav. {\bf 38}, 145002 (2021), arXiv:2103.01140.

\bibitem{RoperPol:2018sap}
A.~Roper~Pol, A.~Brandenburg, T.~Kahniashvili, A.~Kosowsky, and S.~Mandal,
\newblock Geophys. Astrophys. Fluid Dynamics {\bf 114}, 130 (2020),
  arXiv:1807.05479.

\bibitem{DATA}
T.~{Kahniashvili}, E.~{Clarke}, J.~{Stepp}, and A.~{Brandenburg},
\newblock {Datasets for Big bang nucleosynthesis limits and relic gravitational
  waves detection prospects, doi:10.5281/zenodo.5709176 (v2021.11.18)}; see
  also \url{http://www.nordita.org/~brandenb/projects/GWs-BBN/} for easier
  access .

\end{thebibliography}

\clearpage

\noindent
{\Large Supplementary Material to ``Big Bang Nucleosynthesis Limits and Relic Gravitational Waves Detection Prospects"}
\author{Tina~Kahniashvili}
\email{tinatin@andrew.cmu.edu}
\affiliation{McWilliams Center for Cosmology and Department of Physics, Carnegie Mellon University, Pittsburgh, PA 15213, USA}
\affiliation{School of Natural Sciences and Medicine, Ilia State University, 0194 Tbilisi, Georgia}
\affiliation{Abastumani Astrophysical Observatory, Tbilisi, GE-0179, Georgia}

\author{Emma Clarke}
\email{emmaclar@andrew.cmu.edu}
\affiliation{McWilliams Center for Cosmology and Department of Physics, Carnegie Mellon University, Pittsburgh, PA 15213, USA}

\author{Jonathan Stepp}
\email{jdstepp@andrew.cmu.edu}
\affiliation{McWilliams Center for Cosmology and Department of Physics, Carnegie Mellon University, Pittsburgh, PA 15213, USA}

\author{Axel~Brandenburg}
\email{brandenb@nordita.org}
\affiliation{Nordita, KTH Royal Institute of Technology and Stockholm University, 10691 Stockholm, Sweden}
\affiliation{Department of Astronomy, AlbaNova University Center, Stockholm University, 10691 Stockholm, Sweden} 
\affiliation{School of Natural Sciences and Medicine, Ilia State University, 0194 Tbilisi, Georgia}
\affiliation{McWilliams Center for Cosmology and Department of Physics, Carnegie Mellon University, Pittsburgh, PA 15213, USA}
\maketitle

\section{Magnetic Field Bounds}

The bound on extra relativistic degrees of freedom at big bang nucleosynthesis (BBN) can be expressed as
\begin{equation}
    \frac{\rho_B(T_{\rm BBN})}{\rho_{\gamma}(T_{\rm BBN})} = f,
\label{eq:bbnratio}
\end{equation}
where we have assumed that all the extra relativistic energy density
is entirely due to the magnetic energy density $\rho_B$, $\rho_\gamma$
is the energy density in photons, $T_{\rm BBN}$ is the temperature
of helium synthesis
and $f \equiv \frac{7}{8}(\frac{4}{11})^{4/3} \dneff$.

The photon energy density as a function of temperature is $\rho_\gamma = (\pi^2/15)\,T_\gamma^4$. The magnetic energy density is related to the magnetic field strength $B$ as $\rho_B = B^2/ 8\pi$ (in Gaussian units). The magnetic field strength dilutes with the expansion of the universe as $B \sim a^{-2}$ where $a$ is the cosmological scale factor. The comoving magnetic field strength is given by $B^{\rm co} = (a/a_0)^2 B(a)$, where $a_0$ is the scale factor today. Substituting these values into the equation \ref{eq:bbnratio}, the BBN limit on the field strength today is given by
\begin{equation}
    B_*^{\rm co} \leq \bigg(\frac{a_{\rm BBN}}{a_0}\bigg)^2 \sqrt{8\pi f \rho_\gamma(T_{\rm BBN})}.
\end{equation}

Obtaining the ratio of the scale factors via entropy conservation, normalizing such that $a_0$ = 1, the bound is given by
\begin{equation}
    \frac{B_*^{\rm co}}{\rm Gauss} \leq \big(8.06\times10^{-6}\big)\,f^{1/2}\,g_{\rm BBN}^{-2/3}
\end{equation}
where $g_{\rm BBN}$ is the relativistic degrees of freedom at $T_{\rm BBN}$.
There is no explicit dependence on temperature, however, the total
number of relativistic degrees of freedom $g_{\rm BBN}$ does depend on
the temperature.
At $T_{\rm BBN}=$ 0.1 MeV, the temperature at which deuterium synthesis
starts, neutrinos have already decoupled, electrons and positrons have already become nonrelativistic, and
$g_{\rm BBN}(T=0.1{\rm MeV}) \simeq 3.4$. For $\dneff = 0.122$, we find
$f = 0.028$ and the maximum comoving field strength at BBN is $B_{\rm
BBN}^{\rm max} = 6.2 \times 10^{-7}$ G.

\section{Numerical Set-up/Gravitational Waves}

We consider the radiation-dominated epoch at electroweak
(EW) and quantum chromodynamic (QCD) energy scales and
compute the strains $h_+$ and $h_\times$ for the two linear
polarization modes by solving the linearized equation for
gravitational waves (GWs),
\EQ
\frac{\partial^2}{\partial t^2} \tildeh_{+/\times}
+\kk^2\tildeh_{+/\times}={6\over a}\,\tildeT_{+/\times},
\label{d2hdt2}
\EN
where $\tildeT_{+/\times}$ are the $+$ and $\times$ polarizations
of the Fourier transform of the total stress
${\TTT}_{ij}=u_i u_j-B_i B_j$, normalized by the radiation
energy density, with $t$ and $\kk$ the time and wave vector
normalized by the Hubble parameter at the time of generation, and
$\BB=\nab\times\AAA$ and $\uu$ are obtained by solving the
equation for the magnetic vector potential
\EQ
\frac{\partial\AAA}{\partial t}=
\uu\times\BB+\eta\nabla^2\AAA,
\label{dAdt}
\EN
together with \cite{Brandenburg:1996fc}
\begin{equation}
\frac{\partial\uu}{\partial t}=
-\uu\cdot\nab\uu-{1\over4}\nab\ln\rho
+{3\over4\rho}\JJ\times\BB
+\FFF_\nu+\FFF,
\label{dudt}
\end{equation}
\begin{equation}
\frac{\partial\ln\rho}{\partial t}=
-\frac{4}{3}\left(\nab\cdot\uu+\uu\cdot\nab\ln\rho\right) + {\cal H},
\label{dlnrhodt}
\end{equation}
where $\FFF=(\nab\cdot\uu+\uu\cdot\nab\ln\rho)\uu/3 
-[\uu\cdot(\JJ\times\BB)+\JJ^2/\sigma]\uu/\rho$,
and ${\cal H}=[\uu\cdot(\JJ\times\BB)+\JJ^2/\sigma]\rho$
are higher order terms in the Lorentz factor that are retained in the calculation,
and $\FFF_\nu=2\nab\cdot(\rho\nu\SSSS)/\rho$ is the viscous force,
where ${\sf S}_{ij}=\half(u_{i,j}+u_{j,i})-\onethird\delta_{ij}\nab\cdot\uu$
are the components of the rate-of-strain tensor with commas denoting partial
derivatives, and $\nu$ is the kinematic viscosity.
In all cases considered below, we assume a magnetic Prandtl number
of unity, i.e., $\nu/\eta=1$.
In \Tab{Truns}, we summarize the parameters for runs~a--d
and A--D for the QCD and EW energy scales, respectively.
Here, $\hrms^{\rm sat}$ refers to the value of $\braket{h_+^2+h_\times^2}^{1/2}$
evaluated during the final stationary regime.

\begin{table*}[t!]\caption{Summary of the runs.}\vspace{12pt}\centerline{\begin{tabular}{lcccccccc}Run & $f_0$ & $\nu$ & $\EEM^{\rm max}$ & $\EEGW^{\rm sat}$ & $\hrms^{\rm sat}$ & $\Brms$ [$\mu$G] & $h_0^2\OmGW$ & $h_{\rm c}$ \\\hline
F
a & $5\times10^{-1}$ & $2\times10^{-2}$ & $1.40\times10^{-0}$ & $2.6\times10^{-1}$ & $2.7\times10^{-1}$ & $4.7$ & $8.04\times10^{-6}$ & $2.69\times10^{-13}$\\
a2& $3\times10^{-1}$ & $2\times10^{-2}$ & $5.08\times10^{-1}$ & $3.0\times10^{-2}$ & $9.2\times10^{-2}$ & $2.9$ & $9.19\times10^{-7}$ & $9.19\times10^{-14}$\\
b & $3\times10^{-1}$ & $5\times10^{-3}$ & $9.40\times10^{-1}$ & $5.4\times10^{-2}$ & $1.4\times10^{-1}$ & $3.9$ & $1.66\times10^{-6}$ & $1.36\times10^{-13}$\\
c & $2\times10^{-1}$ & $5\times10^{-3}$ & $4.26\times10^{-1}$ & $9.4\times10^{-3}$ & $5.7\times10^{-2}$ & $2.6$ & $2.90\times10^{-7}$ & $5.73\times10^{-14}$\\
d & $1\times10^{-1}$ & $5\times10^{-3}$ & $1.09\times10^{-1}$ & $5.5\times10^{-4}$ & $1.4\times10^{-2}$ & $1.3$ & $1.71\times10^{-8}$ & $1.38\times10^{-14}$\\
\hline
A  &$7\times10^{-3}$ & $5\times10^{-5}$ & $4.05\times10^{-1}$ & $3.0\times10^{-5}$ & $3.1\times10^{-5}$ & $2.5$ & $4.93\times10^{-10}$ & $2.46\times10^{-20}$\\
A' &$7\times10^{-3}$ & $5\times10^{-5}$ & $3.94\times10^{-1}$ & $2.4\times10^{-5}$ & $2.7\times10^{-5}$ & $2.5$ & $3.91\times10^{-10}$ & $2.19\times10^{-20}$\\
A2 &$7\times10^{-3}$ & $1\times10^{-4}$ & $1.91\times10^{-1}$ & $9.5\times10^{-6}$ & $2.0\times10^{-5}$ & $1.8$ & $1.56\times10^{-10}$ & $1.61\times10^{-20}$\\
O1 &$5\times10^{-3}$ & $5\times10^{-5}$ & $1.82\times10^{-1}$ & $5.4\times10^{-6}$ & $1.4\times10^{-5}$ & $1.7$ & $8.86\times10^{-11}$ & $1.12\times10^{-20}$\\
O1'&$5\times10^{-3}$ & $5\times10^{-5}$ & $1.74\times10^{-1}$ & $4.3\times10^{-6}$ & $1.2\times10^{-5}$ & $1.7$ & $7.07\times10^{-11}$ & $9.65\times10^{-21}$\\
O2 &$5\times10^{-3}$ & $1\times10^{-4}$ & $7.50\times10^{-2}$ & $1.7\times10^{-6}$ & $8.4\times10^{-6}$ & $1.1$ & $2.84\times10^{-11}$ & $6.67\times10^{-21}$\\
B  &$2\times10^{-3}$ & $2\times10^{-6}$ & $9.67\times10^{-2}$ & $5.6\times10^{-7}$ & $5.2\times10^{-6}$ & $1.2$ & $9.24\times10^{-12}$ & $4.17\times10^{-21}$\\
C2 &$1\times10^{-3}$ & $2\times10^{-6}$ & $2.74\times10^{-2}$ & $3.1\times10^{-8}$ & $1.3\times10^{-6}$ & $0.66$ & $5.03\times10^{-13}$ & $1.03\times10^{-21}$\\
C  &$1\times10^{-3}$ & $2\times10^{-7}$ & $3.35\times10^{-2}$ & $3.5\times10^{-8}$ & $1.3\times10^{-6}$ & $0.73$ & $5.80\times10^{-13}$ & $1.07\times10^{-21}$\\
D  &$6\times10^{-4}$ & $2\times10^{-7}$ & $1.68\times10^{-2}$ & $5.3\times10^{-9}$ & $7.1\times10^{-7}$ & $0.52$ & $8.73\times10^{-14}$ & $5.64\times10^{-22}$\\
\label{Truns}
\end{tabular}}
\end{table*}

As in Ref.~\cite{Kahniashvili:2020jgm}, hereafter K+21, we compute GW
generation from magnetically driven turbulence.
The driving is applied during the time interval $1\leq t\leq2$, where $t$
is the conformal time.
As in K+21, we then decrease the driving linearly in time until $t=3$,
when the driving is turned off completely.
We perform series of runs where we vary the strength of the forcing $f_0$
and keep the viscosity $\nu$ unchanged.
However, it is not possible to explore the regime of strong magnetic
energy at the same small values of $\nu$ that we were able to use for
smaller magnetic energies.
This is because for strong magnetic fields, the turbulence becomes more
intense and more viscosity is needed to dissipate all this energy at
the finite numerical resolution available.

\begin{figure}[ht!]\begin{center}
\includegraphics[width=\columnwidth]{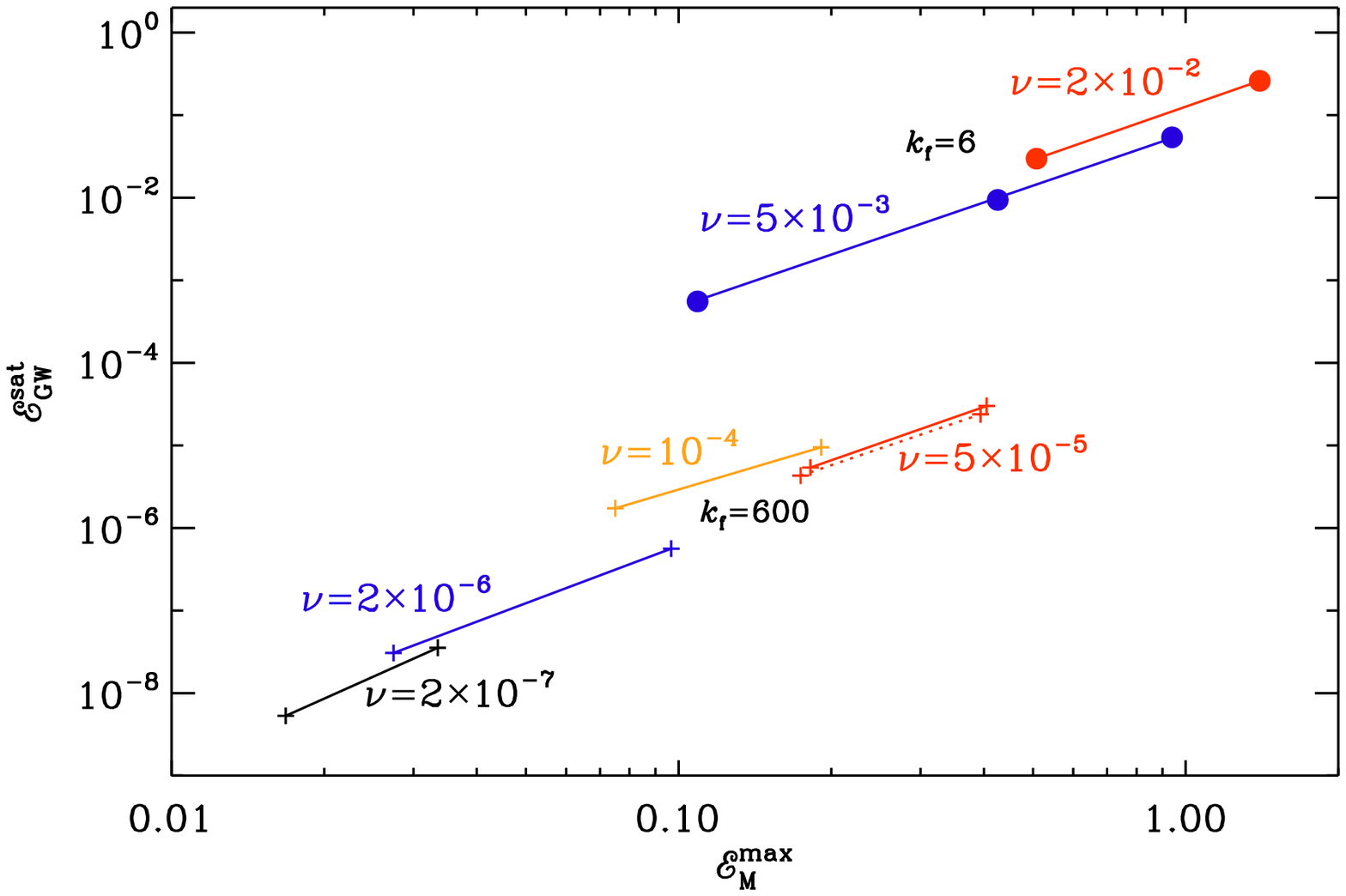}
\end{center}\caption[]{
Dependence of $\EEGW^{\rm sat}$ on $\EEM^{\rm max}$ for magnetically
driven turbulence at different forcing strengths and viscosities
for $\kf=6$ (upper red and blue lines) and 
$\kf=600$ (lower red, orange, blue, and black lines).
The red dashed line for $\kf=600$ denotes runs where the driving
is turned off abruptly at $t=2$.
}\label{ppe}\end{figure}

\begin{figure}[ht!]\begin{center}
\includegraphics[width=\columnwidth]{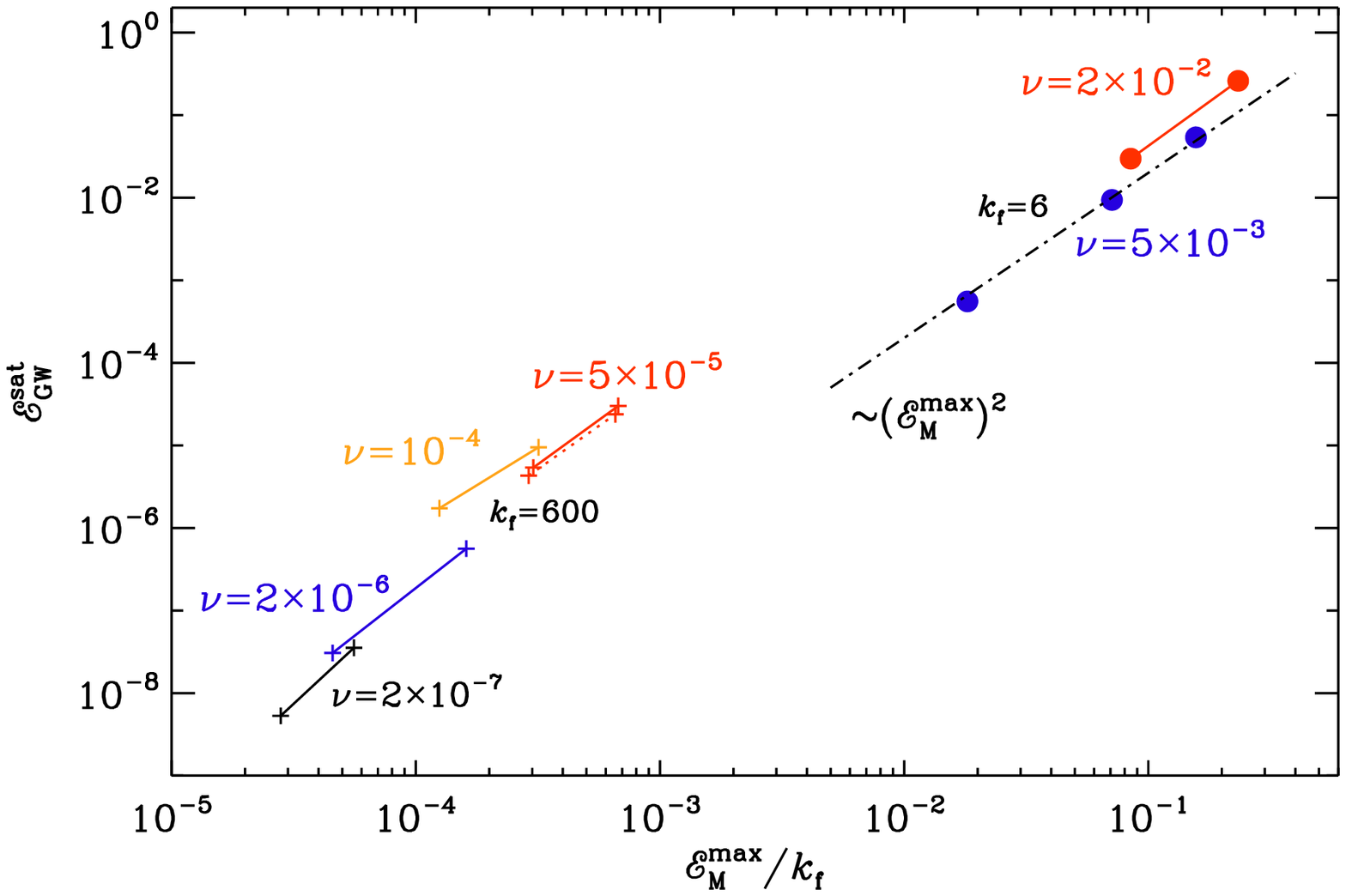}
\end{center}\caption[]{
Dependence of $\EEGW^{\rm sat}$ on $\EEM^{\rm max}/\kf$ for the
same runs as in \Fig{ppe}.
}\label{ppe_scl}\end{figure}

In \Fig{ppe}, we show the resulting dependence of the GW energy $\EEGW$
on the magnetic energy $\EEM$ for six sets of runs with fixed viscosity,
different forcing strengths, and different forcing wavenumbers,
corresponding to the runs denoted with labels a--d, A--D, and O.
In all cases, we take the magnetic Prandtl number to be unity, i.e.,
the magnetic diffusivity is set equal to the value of $\nu$.
We also compare with several other sets of runs where we change the
forcing.

\begin{figure*}[t!]\begin{center}
\includegraphics[width=\textwidth]{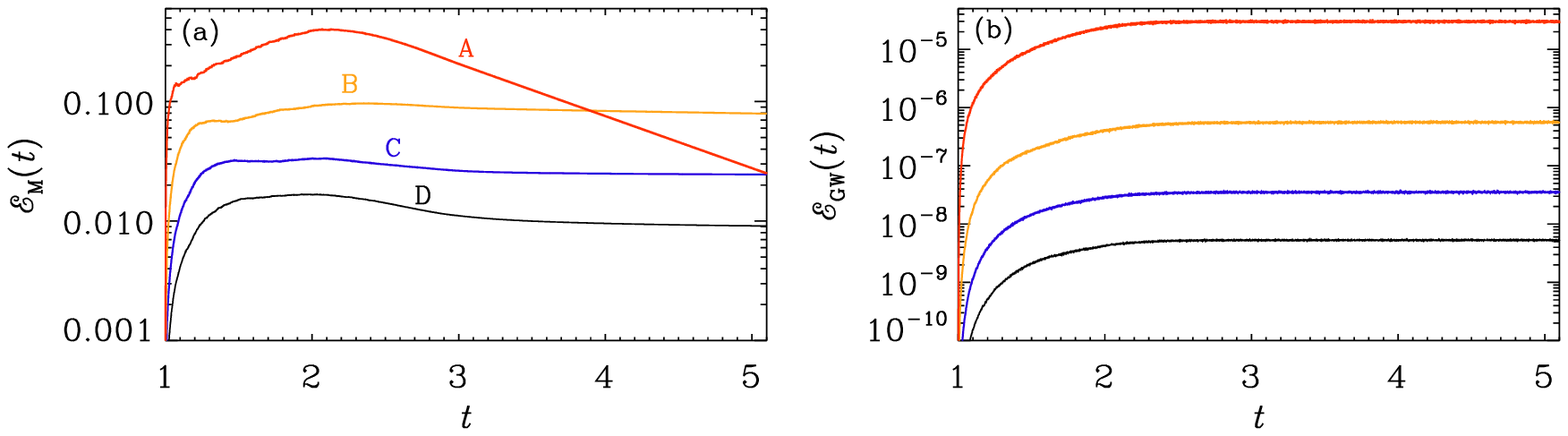}
\end{center}\caption[]{
Evolution of (a) $\EEM(t)$ and (b) $\EEGW(t)$ for Runs~A--D of \Tab{Truns}.
Note the rapid decay for Run~A with the largest viscosity.
}\label{pcomp_M512k2_large}\end{figure*}

In \Tab{Truns}, we summarize the parameters for four runs (A--D), which
correspond to the less viscous ones for each of the four pairs shown
in \Fig{ppe}.
One exception is Run~D, which has the same viscosity as Run~C and is
denoted in \Fig{ppe} by a red line.
Run~D is the same one as Run~M1 of K+21.
The values of $\EEM$ and $\EEGW$ agree with those of K+21 for this run,
but those of $\hrms$ are here a bit smaller.
In fact, a closer inspection of the time series of $\hrms(t)$ revealed
that it reaches a steady state much later than $\EEGW(t)$.
Therefore, averaging can begin only later than for $\EEGW$.
Since $\hrms$ is found to decrease somewhat after having reached
a maximum, the new value in \Tab{Truns} is now about 20\%
smaller than that given in K+21.

The data for $\EEGW$ follow a power law scaling, $\propto\EEM^n$,
where $n=2.7$ for the points with the smallest viscosity.
This is steeper than the quadratic scaling found in the work of
\cite{RoperPol:2019wvy},
where the driving was applied for a much shorter time
interval, $1\leq t\leq1.1$.
Furthermore, for fixed values of $\nu$, we find smaller local values of
$n$, at least for the larger magnetic energies shown in \Fig{ppe}.
We also checked that these scalings are not significantly
affected if the driving was turned off abruptly after $t=2$.
This is shown as the dotted line in \Fig{ppe}
for $\nu=5\times10^{-5}$.

Comparing the lines for $\nu=5\times10^{-5}$ and $\nu=10^{-4}$
in \Figs{ppe}{ppe_scl},
we see that the decline of $\EEM$ is stronger than that of $\EEGW$.
This suggests that $\EEM$ suffers more strongly from
the increase of viscosity and magnetic diffusivity, and that
$\EEGW$ is less sensitive to the change of $\nu$.
However, one has to remember that GWs are solely the result of the
magnetic and hydrodynamic stresses.
One sees that the runs with smaller values of $\nu$ all have a faster
rise of $\EEM(t)$ early on, which also translates into a rapid increase
of $\EEGW(t)$.
It is unclear, however, whether this aspect of the model with applied
magnetic driving is realistic and whether this would also be borne out
by a more physical implementation of a magnetogenesis model.

Next, we show in \Fig{pcomp_M512k2_large} the evolution of $\EEM(t)$
and $\EEGW(t)$ with time.
We see that for Runs~C and D, $\EEM$ has reached a plateau well before
$t=2$, while for Run~A, a maximum is reached only at $t=2$, i.e., the
time when the driving is decreased.
Moreover, for Run~A, there is a strong temporal decline of magnetic
energy due to strong viscous damping.
Nevertheless, similar GW energies are obtained in this case.
The value of $\EEGW=3\times10^{-5}$ given in \Tab{Truns}
corresponds to $h_0^2\OmGW=4.93\times10^{-10}$, which is four
orders of magnitude larger than for Run~D.


\end{document}